\documentclass{appolb}
\usepackage{epsfig}

\newcommand{\bea}{\begin{eqnarray}} 
\newcommand{\eea}{\end{eqnarray}}

\begin{document}
\title{FSR at leading and next-to-leading order  \\
 in the radiative return at meson factories 
\thanks{Presented by H. Czy{\.z}    
 at XXVII International Conference of Theoretical Physics,
 `Matter To The Deepest', Ustro{\'n}, 15-21 September 2003, Poland.\\
 Work supported in part by 
 EC 5-th Framework Program under contracts HPRN-CT-2000-00149,
  and HPRN-CT-2002-00311 (EURIDICE network) and
  Polish State Committee for Scientific Research
  (KBN) under contract 2 P03B 017 24.
}
}
\author{Henryk Czy\.z, Agnieszka Grzeli{\'n}ska
\address{Institute of Physics, University of Silesia,
PL-40007 Katowice, Poland.}
}
\maketitle
\vspace{-4.5 cm}
\hfill{\bf TTP03-031}
\vspace{+4.5cm}

\begin{abstract}
  The impact of final-state radiation (FSR) on the radiative
 return method for the extraction of the $e^+e^-$ hadronic cross section
 is discussed in detail. Possible experimental tests of the model
 dependence of FSR are proposed for the $\pi^+\pi^-$ hadronic final state.
  The importance of the $\pi^+\pi^-\gamma$ final state contribution
 to the muon anomalous magnetic moment is investigated,
 and a method based on the radiative return is proposed 
 to extract these contributions from data. 
\end{abstract}
  
\section{Introduction to the radiative return method }
 Electron--positron annihilation into hadrons in the low energy region
   is crucial
 for predictions of the hadronic contributions to $a_\mu$, the anomalous 
 magnetic moment of the muon, and to the running of the electromagnetic
 coupling from its value at low energy up to $M_Z$ 
(For reviews see 
 e.g.  [1-7]
 the most recent experimental result for $a_\mu$ 
  is presented in \cite{Bennet}).
 Measurements of the hadronic cross section for electron--positron annihilation
 were traditionally performed  by varying the 
 beam energy of the collider. The 
 \({\mathrm  \Phi}\)- and B-meson factories
 allow to use the radiative return to explore the whole energy
 region from threshold up to the energy of the collider. 
 Even if the photon radiation 
 from the initial state reduces the cross section by a factor
 ${\cal O}(\alpha/\pi)$, this is easily compensated by the 
 enormous luminosity of these `factories'. A number of experimental results
 based on the radiative return
 was already published [9-18]
 and in the near future one can expect
 much more data covering large variety of hadronic final states.

 The radiative return method \cite{Binner:1999bt} (see also \cite{Zerwas}),
  relies on the following factorisation property of the cross section

\bea
 \frac{d\sigma}{dQ^2d\Omega_\gamma}
 \left(e^+e^- \to \mathrm{hadrons} + \gamma\right) = 
H(Q^2,\Omega_\gamma) \ \sigma(e^+e^-\to \mathrm{hadrons,Q^2}) \ ,
 \nonumber \\
 \label{rr}
\eea

\noindent
 where $Q^2$ is the invariant mass of the hadronic system, $\Omega_\gamma$ 
 denotes the photon polar and azimuthal angles, and the function
 $H(Q^2,\Omega_\gamma)$ is given by QED lepton-photon interactions,
 thus known in principle with any required precision.
 The formula (\ref{rr}) is valid for 
 a photon emitted from initial state leptons (ISR)
 and what is more important
 similar factorisation formula applies for the emission  
 of an arbitrary number
 of photons or even lepton pairs \cite{KKKS88} from initial state leptons. 
 Let's forget for a while about FSR.
 In that case
 one can, by
 measuring the $Q^2$ differential cross section
 of the process 
 $e^+e^- \to \mathrm{hadrons} + \mathrm{photons} +
 (possibly)\ \mathrm{lepton \ pairs}$ (called $\sigma_{RR}$ from here on)
  and knowing function the
 $H(Q^2,...)$, extract the value of $\sigma(e^+e^- \to \mathrm{hadrons})$.
 The $...$ in $H(Q^2,...)$ stand for the phase space variables of
 photons and/or lepton pairs. 
 However, in practice the cross section $\sigma_{RR}$ is measured within 
 a given experimental setup, which corresponds to an integral  
 over a complicated phase space of photons and/or lepton pairs
 and
 thus the use of  Monte Carlo event generators for extraction of 
 the $\sigma(e^+e^-\to \mathrm{hadrons,Q^2})$ become 
 indispensable. Such Monte Carlo programs 
 (EVA \cite{Binner:1999bt,Czyz:2000wh}, 
 PHOKHARA \cite{Rodrigo:2001kf,Czyz:2002np,Czyz:PH03}), were and are being
 developed. The analysis presented
 in this paper is based on the results obtained by means of the program
 PHOKHARA 3.0 \cite{Czyz:PH03} (For further extensive discussions
 of various aspects of the radiative return method not covered by this
 article see [26-31],while for related discussion of the scan method
 look \cite{Gluza:rad,Gluza:eta}).
 
 Further complication arises as the photons and/or lepton pairs are 
 emitted also from final state charged hadrons and special care
 has to be taken when using the radiative return method.
 That problem is discussed extensively in Sections \ref{secFSRLO} 
 and \ref{secFSRNLO}, while Section \ref{gm2} 
 is devoted to NLO contributions
 to $a_{\mu}$.

\section{Leading order FSR}{\label{secFSRLO}}

\begin{figure}[ht]
\begin{center}
\epsfig{file=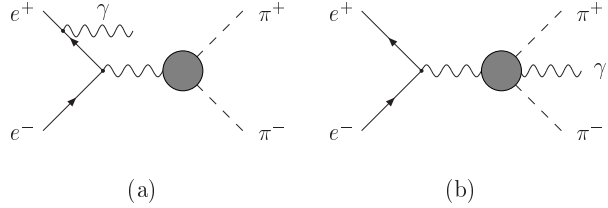,width=8.5cm} 
\vskip -0.2cm
\caption{Leading order contributions to the reaction
$e^+e^-\to\pi^+\pi^-\gamma$ from ISR (a) and FSR (b).}
\label{fig1}
\end{center}
\end{figure}

 Leading order (LO) contributions to the process $e^+e^-\to\pi^+\pi^-\gamma$
 are schematically (permutations omitted) shown in Fig.\ref{fig1}.
 The interference of  ISR, which leads to a C-odd 
(C stands for charge conjugation) configuration of $\pi^+\pi^-$ pair 
 (Fig.1a), with the FSR amplitude from Fig.1b, corresponding to C-even
 $\pi^+\pi^-$ configuration,
 vanishes if a charge symmetric event selection is used.  
 It gives rise, however, to charge asymmetries and charge induced 
 forward--backward asymmetries.

\begin{figure}[ht]
\begin{center}
\epsfig{file=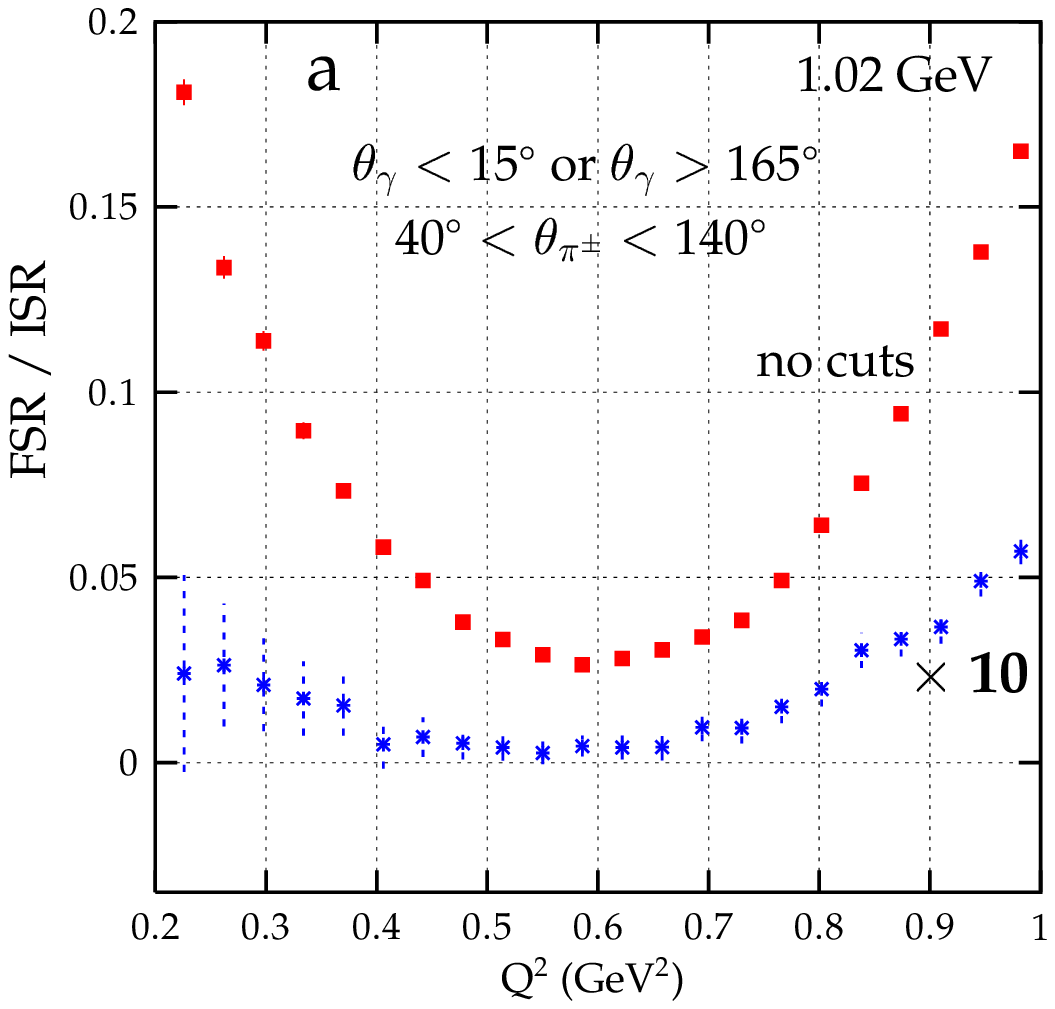,width=5.9cm,height=5.5cm}
\epsfig{file=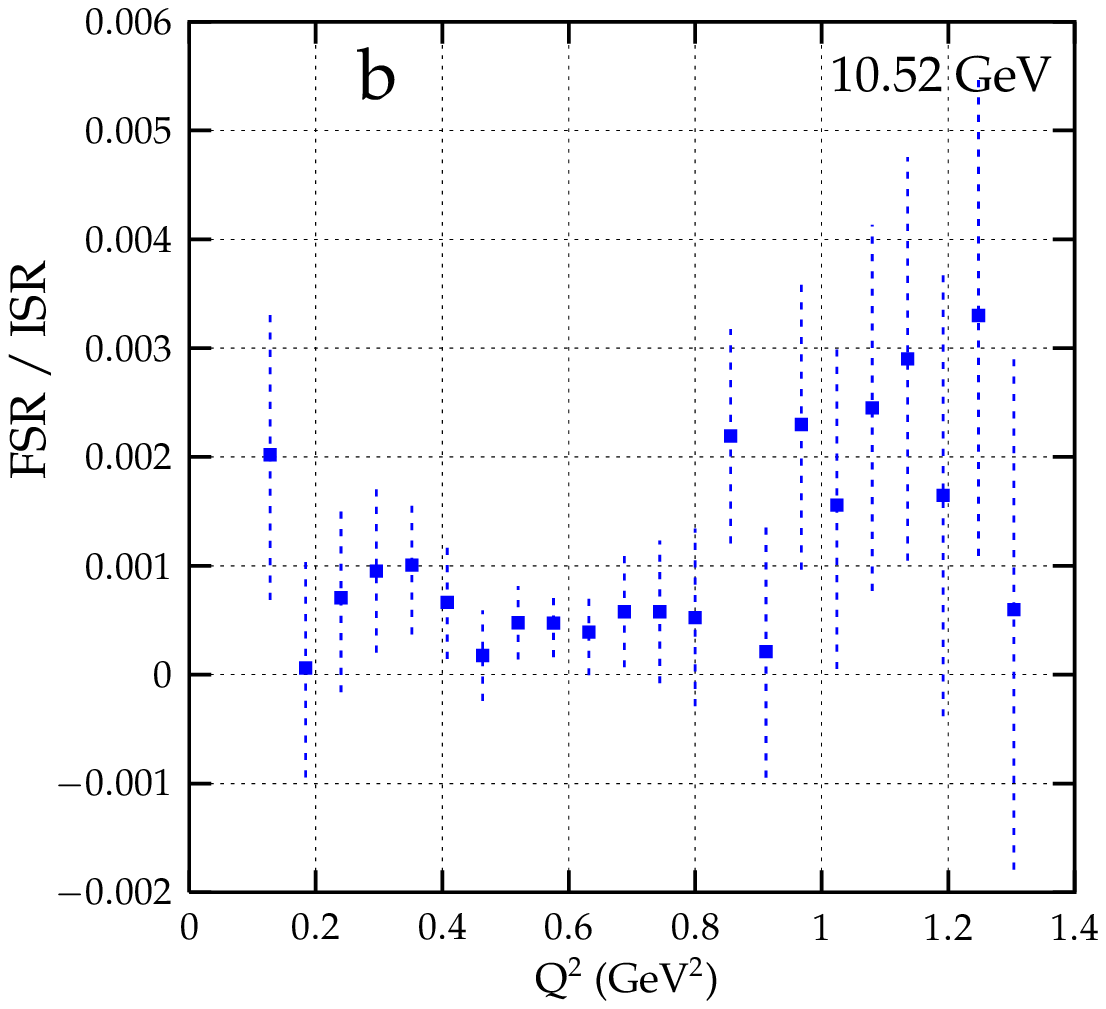,width=5.9cm,height=5.5cm}
\caption{Relative contribution of FSR with respect to ISR
 to the inclusive photon spectrum at \(\sqrt{s}~=\) 1.02 GeV
 (without and with cuts (multiplied by a factor 10)) (a)
  and  \(\sqrt{s}~=\) 10.52 GeV (b).}
\label{fig4}
\end{center}
\end{figure}

The FSR contribution itself can be as big as 20\% of the ISR at
 $\sqrt{s}~=$ 1.02 GeV if no cuts are applied (Fig.\ref{fig4}a). At  
\(\sqrt{s}~=\) 10.52 GeV however, a very energetic photon 
 has to be radiated to produce a pair of charged pions with
 invariant mass around the $\rho$ resonance,
 and its momentum has to be compensated by the
 momentum of the charged pions. As a result, photon and pions
 are produced back to back, thus leading to a negligible contribution
 from LO FSR (Fig.\ref{fig4}b). At DA$\Phi$NE energy 
 FSR has to be controlled through suitable event selection, 
 and an example of a possible event selection
 is shown in Fig.\ref{fig4}a  (lower curve). 
\begin{figure}[ht]
\vskip -0.2 cm
\begin{center}
\epsfig{file=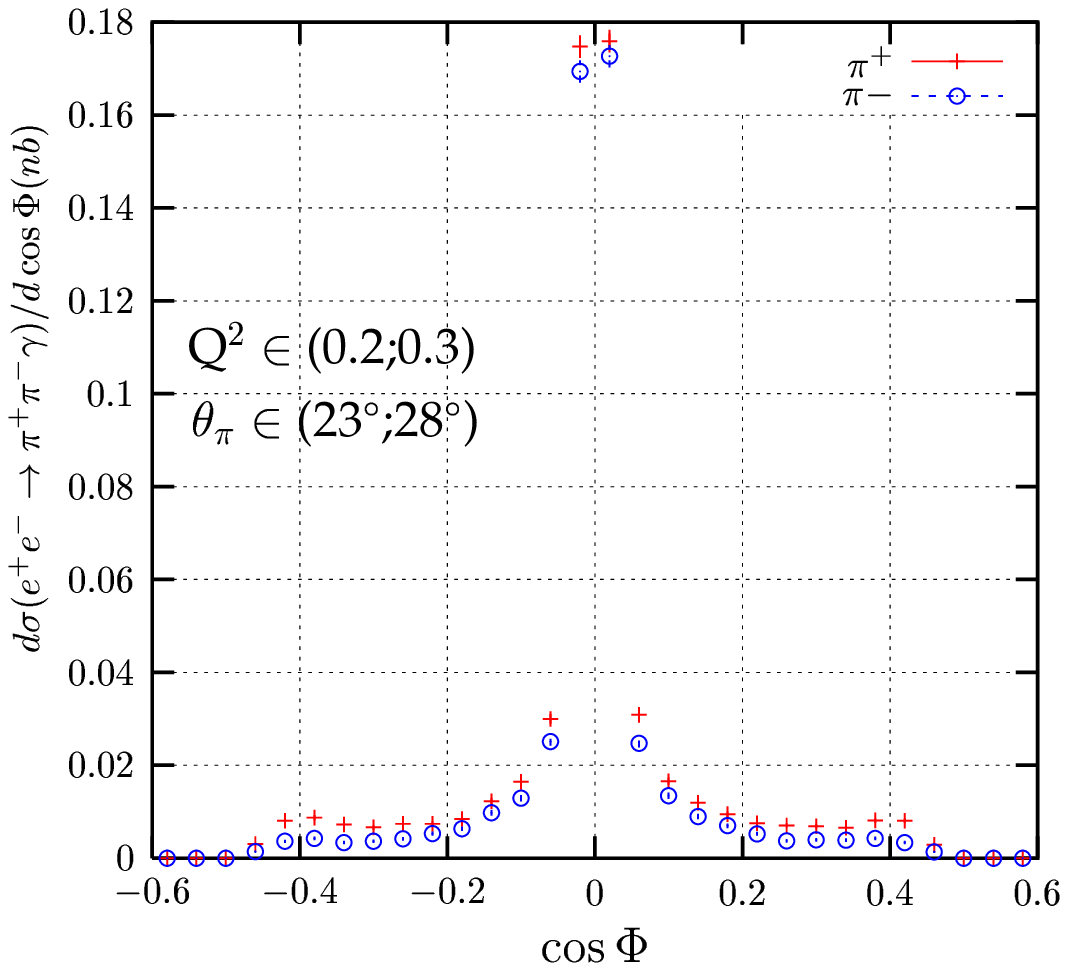,width=5.9cm,height=5.5cm}
\epsfig{file=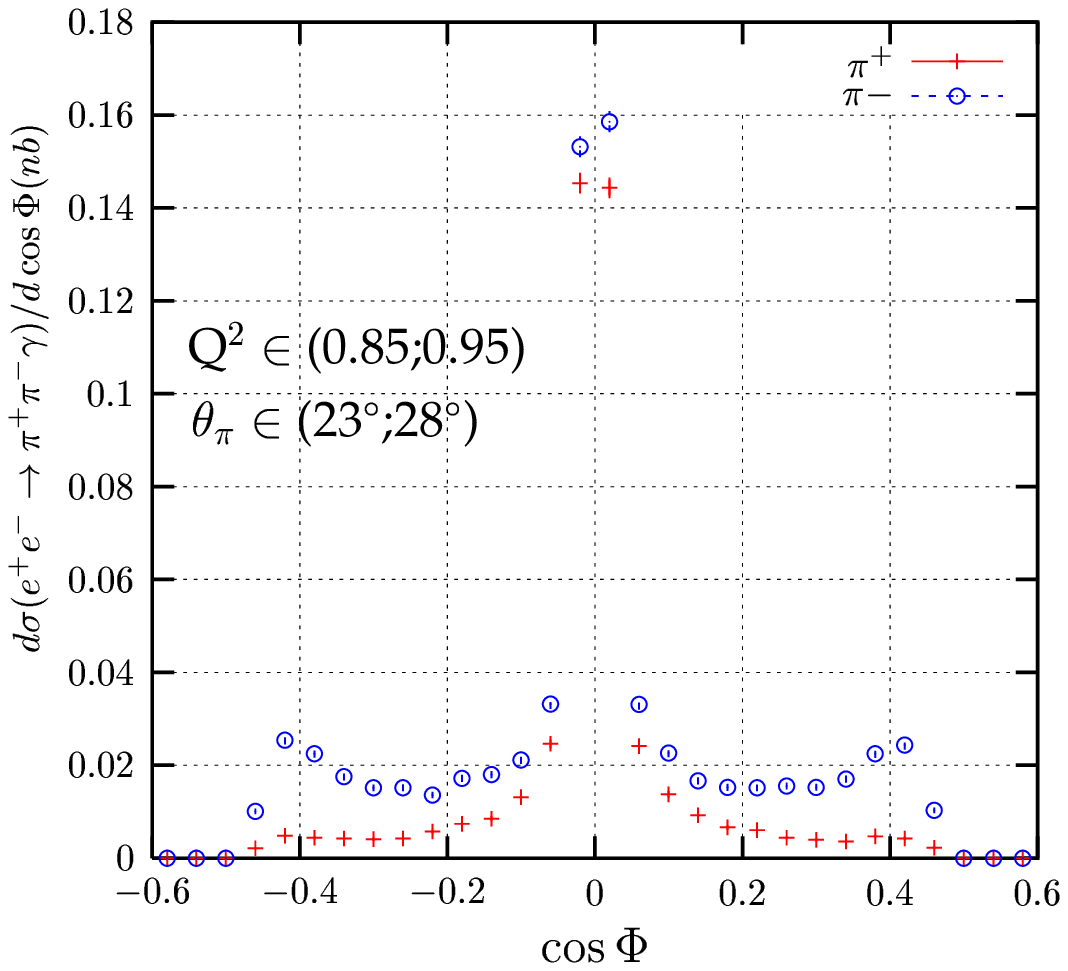,width=5.9cm,height=5.5cm}
\caption{Charge asymmetric distributions of the pions in $\Phi$ (the angle
 between normal to the production plain and the direction 
 of the initial positron) for different values of $Q^2$ 
 and fixed pion polar angles.}
\vskip -0.3 cm
\label{fig5}
\end{center}
\end{figure}

 One can easily reduce the FSR contribution
 to less than 1\% and, relying on a MC generator, subtract it from the
 measured cross section, thus being able to apply the procedure of the
 hadronic cross section extraction described in the previous section.
  The FSR contribution is however model dependent
 and one needs an independent experimental check on the accuracy of the
 model used. It can be done
 just by relaxing the cuts and measuring various charge asymmetric
 distributions. As the actual contribution of FSR to the radiative
 return cross section,  
 is of the order of 1\%, a modest 10\% accuracy of the model will lead to
 an error of 0.1\%, sufficient for any high precision measurement.
 Some of the tests of the model used in EVA and PHOKHARA for FSR
 (point-like pions and scalar QED (sQED)), proposed in \cite{Binner:1999bt},
  were already done by KLOE \cite{Aloisio:2001xq}, where it was
  shown that the charge asymmetry
\begin{equation}
 A(\theta) = \frac{N^{\pi^+}(\theta)-N^{\pi^-}(\theta)}
{N^{\pi^+}(\theta)+N^{\pi^-}(\theta)} \ \ ,
 \label{asym}
\end{equation}
 agrees well with the EVA MC \cite{Binner:1999bt}.
 However additional tests are needed to assure the accuracy
 of the model at the required level
 and comparisons of various charge asymmetric distributions
 between experimental data and MC are indispensable.
 If only pions four-momenta are measured, as done in the KLOE experiment
 at the moment (see  \cite{Achim:radcor02,KLOE:2003}), one arrives at
  distributions as shown in Fig.\ref{fig5}. With
 500 pb$^{-1}$, collected till now by
 KLOE, the 0.1 nb/bin in the plot
 corresponds to 2000 events per bin. Thus that kind of measurement is 
 feasible and tests can be done with the required precision, provided
 systematic errors are small enough. The nontrivial $Q^2$ and polar angle
 dependence of $\Phi$--distributions provides  profound
 cross checks of the tested model.

\section{Next-to-leading order
  hadronic contributions to muon $a_\mu$}{\label{gm2}}

\begin{figure}[ht]
\begin{center}
\epsfig{file=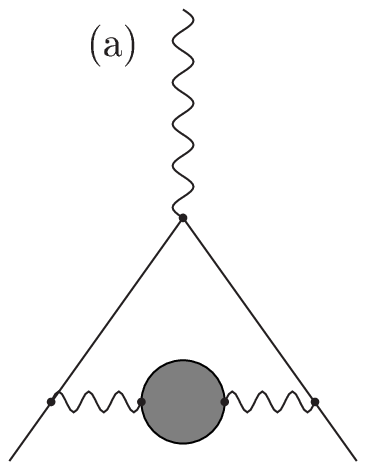,width=4.9cm,height=3.5cm}
\epsfig{file=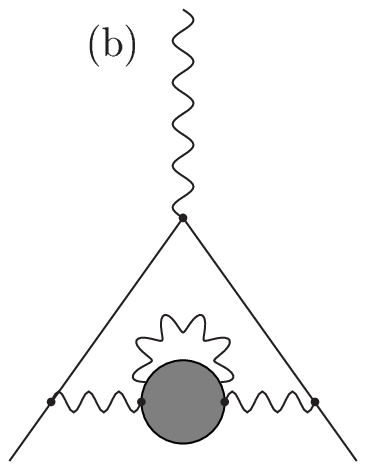,width=4.9cm,height=3.5cm}
\end{center}
\vspace{-1 cm}
\begin{center}
\caption{Hadronic leading order (a)
 and next-to-leading order (hadrons+photon) (b) 
 \hskip 5cm contributions to the muon anomalous 
 magnetic moment.}
\vskip -0.6cm
\label{fig6}
\end{center}
\end{figure}
 The only reliable method of calculation of the hadronic contribution
 to the muon anomalous magnetic moment $a_{\mu}$,
  available till now in the low energy region,
  is based on dispersion relations, where the measured
  $e^+e^-$ hadronic cross section
 is convoluted with the known kernel function (for definitions
 and recent reviews see \cite{Jegetc.,Davier:2003,Nyffeler:2003}).
  The LO hadronic contribution to $a_{\mu}$ enters 
 through the graph(s) shown in Fig.\ref{fig6}a.
 One class of contributions at the next-to-leading order
  (NLO) is shown in Fig.\ref{fig6}b,
   where an additional photon is attached to charged hadron line(s).
 For the two pion plus photon 
 final state one can try to estimate the contribution
 using point-like pions and sQED or just by inserting a quark loop with
 photonic corrections.  
 As shown in \cite{Czyz:PH03},
  one cannot rely on a theoretical model,
 as the values obtained vary very much with the chosen model
 and do depend very strongly on the value of the quark mass.
 The values obtained 
  are:
$\delta a_\mu({\rm quark},\gamma,m_q=180~{\rm MeV})
 = \ 1.880 \times 10^{-10}$,
$\delta a_\mu({\rm quark},\gamma,m_q=66~{\rm MeV})
 = \ 8.577 \times 10^{-10}$ and 
 $\delta a_\mu(\pi^+\pi^-,\gamma) = \ 4.309 \times 10^{-10}$, where
 $m_q=180~{\rm MeV}$ describes well the LO contributions to $a_{\mu}$ , while
 $m_q=66~{\rm MeV}$ describes well the lowest order contribution
 to $\alpha(M_Z)$.

 As the values are of the order of the present error of the hadronic
 contribution to $a_{\mu}$ \cite{Davier:2003} it is important to measure
 also the relevant parts of the
 $e^+e^-\to \pi^+\pi^- \gamma$ cross section and then,
 using dispersion relations, get the contribution of the $\pi^+\pi^- \gamma$
 intermediate state to $a_{\mu}$.

\begin{figure}[ht]
\epsfig{file=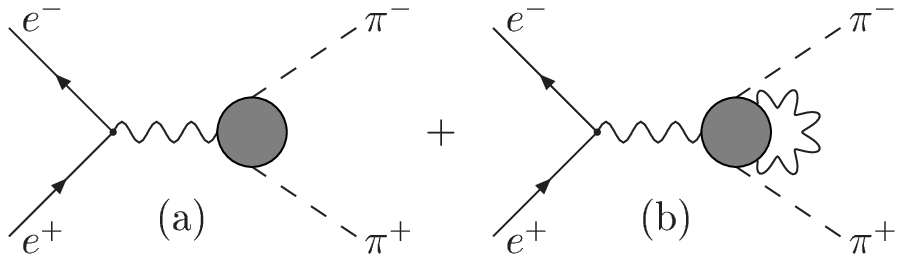,width=7.9cm,height=3.5cm}\hskip 0.6 cm
\epsfig{file=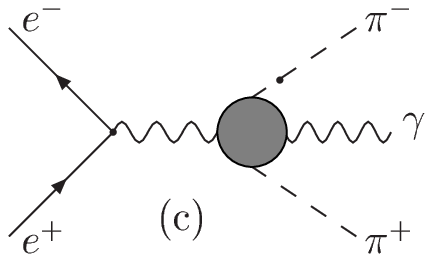,width=3.9cm,height=3.5cm}
\begin{center}
\caption{Leading order (a) and $\pi^+\pi^-\gamma^*$ vertex corrections (b)
 contributions to the reaction $e^+e^-\to\pi^+\pi^-$;
 leading order FSR contribution to the reaction
$e^+e^-\to\pi^+\pi^-\gamma$ (c). }
\label{fig7}
\end{center}
\vspace{-1cm}
\end{figure}

\begin{figure}[hb]
\begin{center}
\vspace{-0.5cm}
\epsfig{file=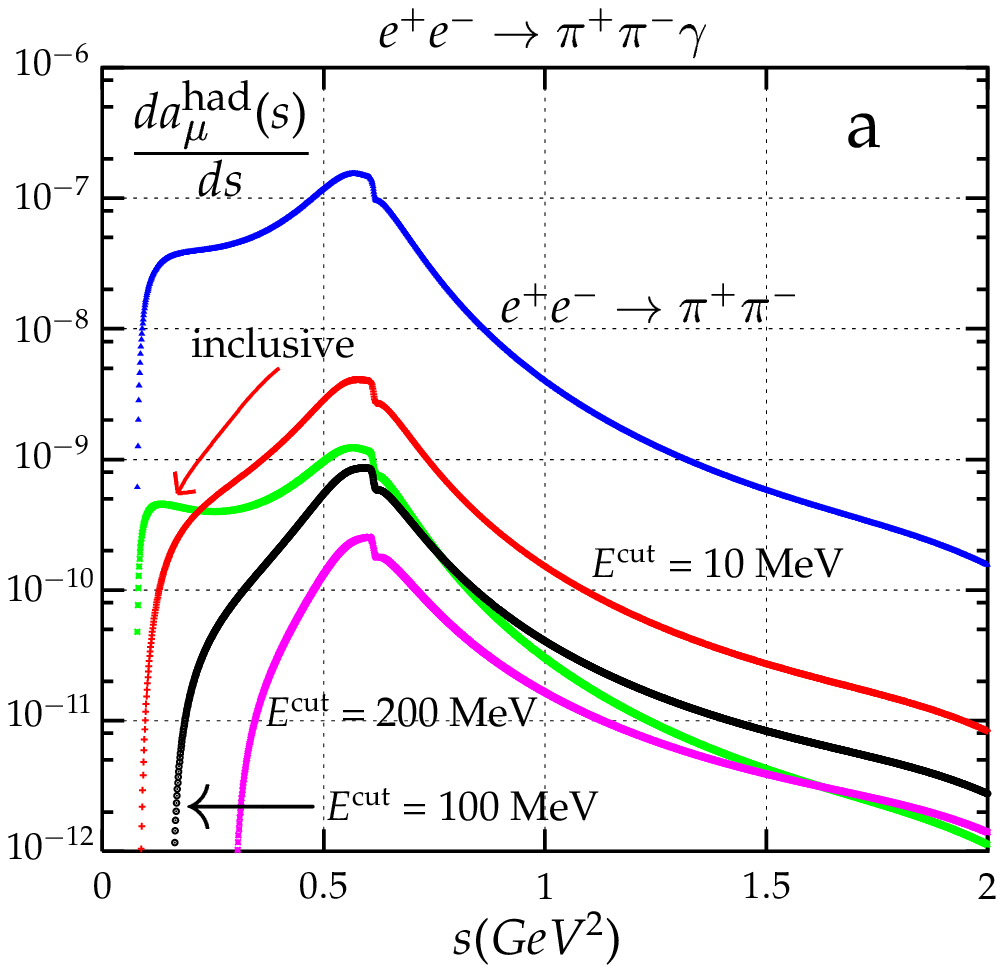,width=6.0cm,height=5.5cm}
\epsfig{file=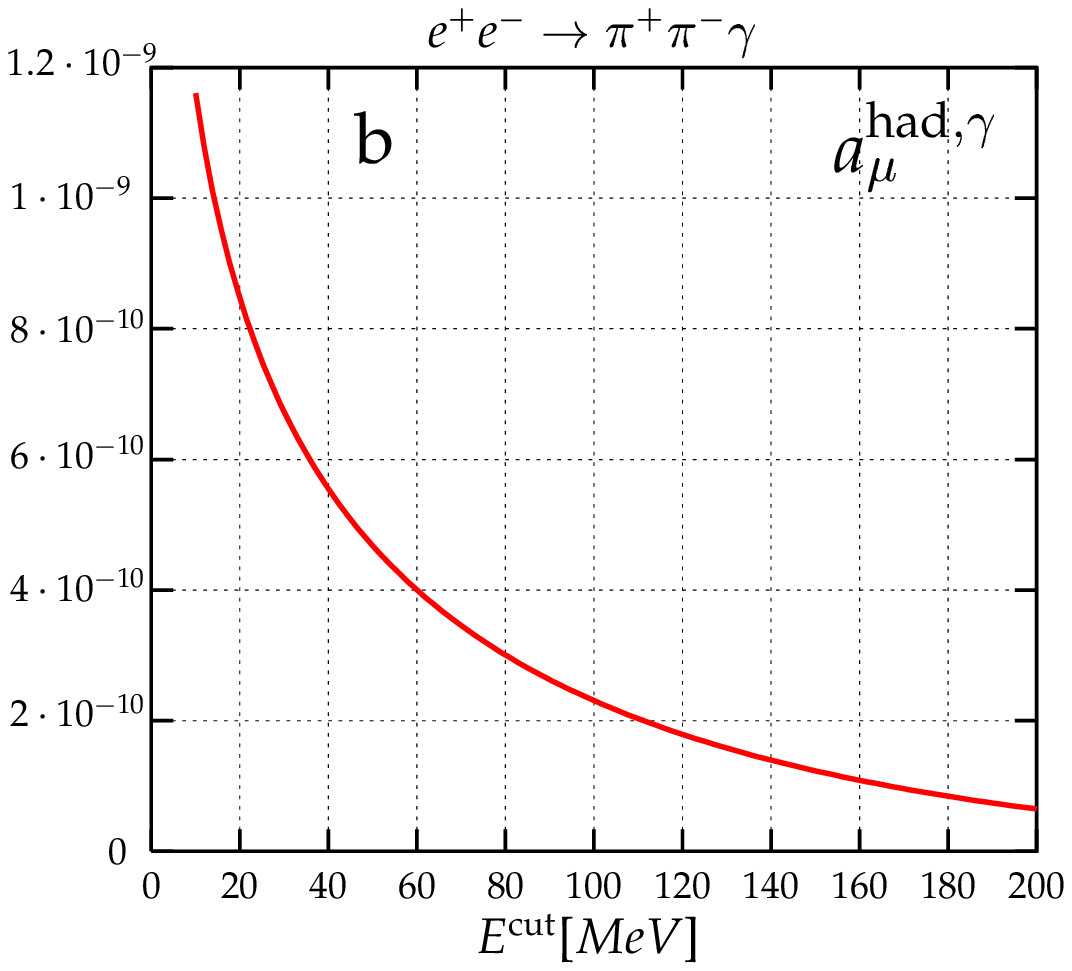,width=6.0cm,height=5.5cm}
\caption{Differential contribution to $a_\mu^\mathrm{had,\gamma}$ 
from $\pi^+\pi^-\gamma$ intermediate states for 
different cutoff values
 compared with complete contribution 
(virtual plus real corrections, labelled `inclusive')
 evaluated in scalar QED (FSR) as well as 
 with contribution from $\pi^+\pi^-$ intermediate state (a) and
 integrated contribution to $a_\mu^\mathrm{had,\gamma}$  
as function of the cutoff $E^\mathrm{cut}$ (b).}
\label{fig8}
\end{center}
\end{figure}

  This is by no means easy
 (for more extensive discussions see  \cite{Czyz:PH03}),
 as the two body cuts of the diagram in Fig.\ref{fig6}b correspond
 to the radiative corrections to the $\pi\pi\gamma$ vertex 
 (Fig.\ref{fig7}b), while the three body cuts correspond to a part of the
 leading order $e^+e^-\to\pi^+\pi^-\gamma$ process (Fig.\ref{fig7}c), with
 the photon emitted from the final states only.
 In practice, when measuring the cross section $\sigma(e^+e^-\to\pi^+\pi^-)$,
 one always measures in combination the contributions from Fig.\ref{fig7}a and
 \ref{fig7}b plus the `soft' part of the contribution from Fig.\ref{fig7}c,
  with the latter
  depending on the actual experimental setup.

The virtual plus soft corrections 
 to the $\pi^+\pi^-\gamma^*$ vertex are negative.
 As a result, the actual contribution from the 
 'hard' part (photon with energy above 
 $E_{\mathrm cut}$) can be larger than the inclusive 
 sum of 'virtual'+'soft'+'hard'  contributions.
  This is shown in Fig.\ref{fig8}a, where
 differential contributions to $a_\mu^\mathrm{had,\gamma}$ from
 $\pi^+\pi^-\gamma$ intermediate states are
 compared with the contribution from $\pi^+\pi^-$.
 The contribution to $a_\mu$, integrated over the whole s--spectrum,
 is shown in Fig.\ref{fig8}b. Thus a special care
 has to be taken when imposing cuts on the photon energy below 50 MeV,
 as it might lead to significant shift for $a_\mu$.

\section{FSR at NLO and tests of the FSR models
 }{\label{secFSRNLO}}

\begin{figure}[ht]
\begin{center}
\hskip +3cm\epsfig{file=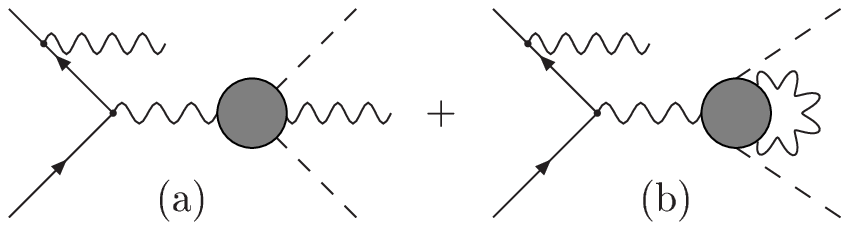,width=8.5cm,height=3.5cm} 
\caption{NLO contributions to the reaction
$e^+e^-\to\pi^+\pi^-\gamma$ from real (both soft and hard)
 FSR emission ($\Delta$IFSNLO(S+H))
  (a) and virtual corrections
 to the $\pi^+\pi^-\gamma^*$ vertex ($\Delta$IFSNLO(V)) (b).}
\label{fig9}
\end{center}
\end{figure}

 The upgrade of the PHOKHARA event generator to version 3.0 
\cite{Czyz:PH03} consisted in adding the diagrams from Fig.\ref{fig9},
 where the photon emitted from the initial state is assumed to be 'hard',
 say with energy above 10 MeV at DA$\Phi$NE and 100 MeV
 at B--factories.

\begin{figure}[ht]
\begin{center}
\epsfig{file=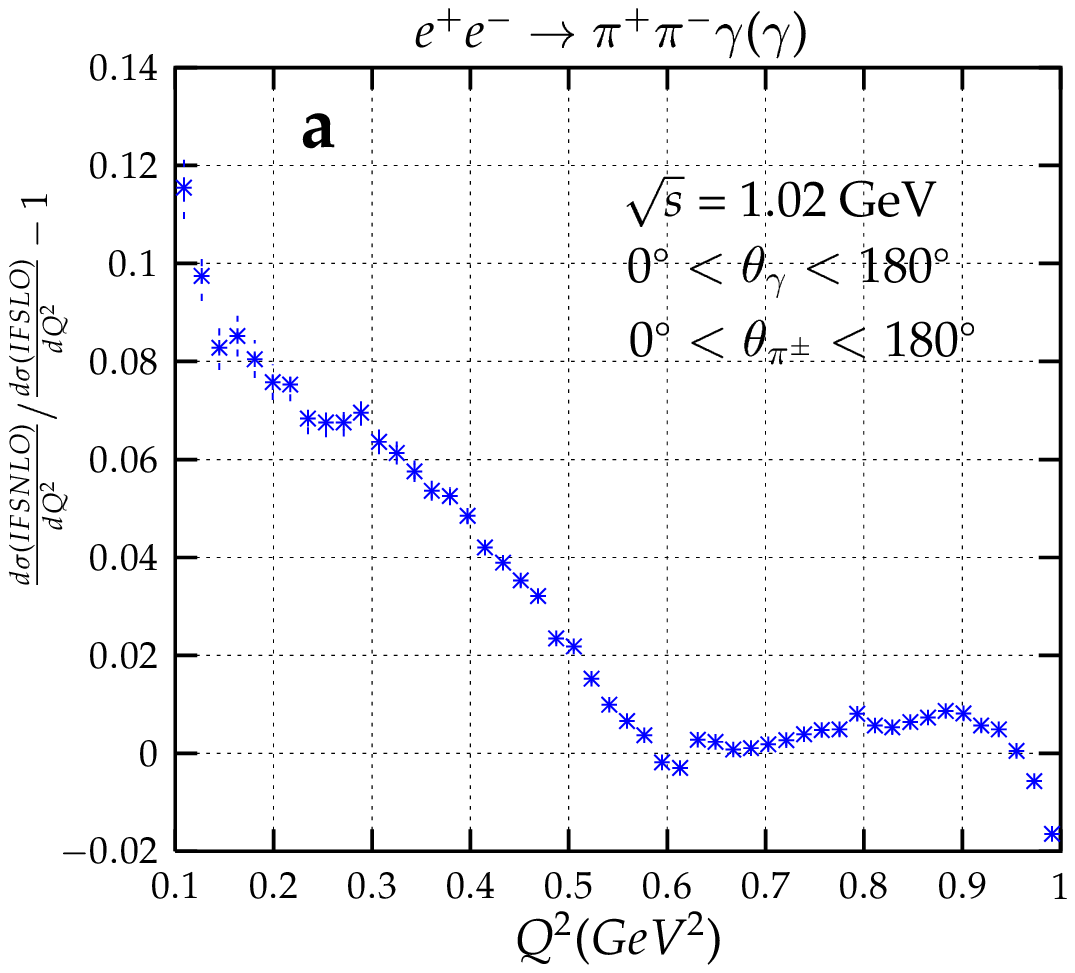,width=6.2cm,height=5.5cm}
\epsfig{file=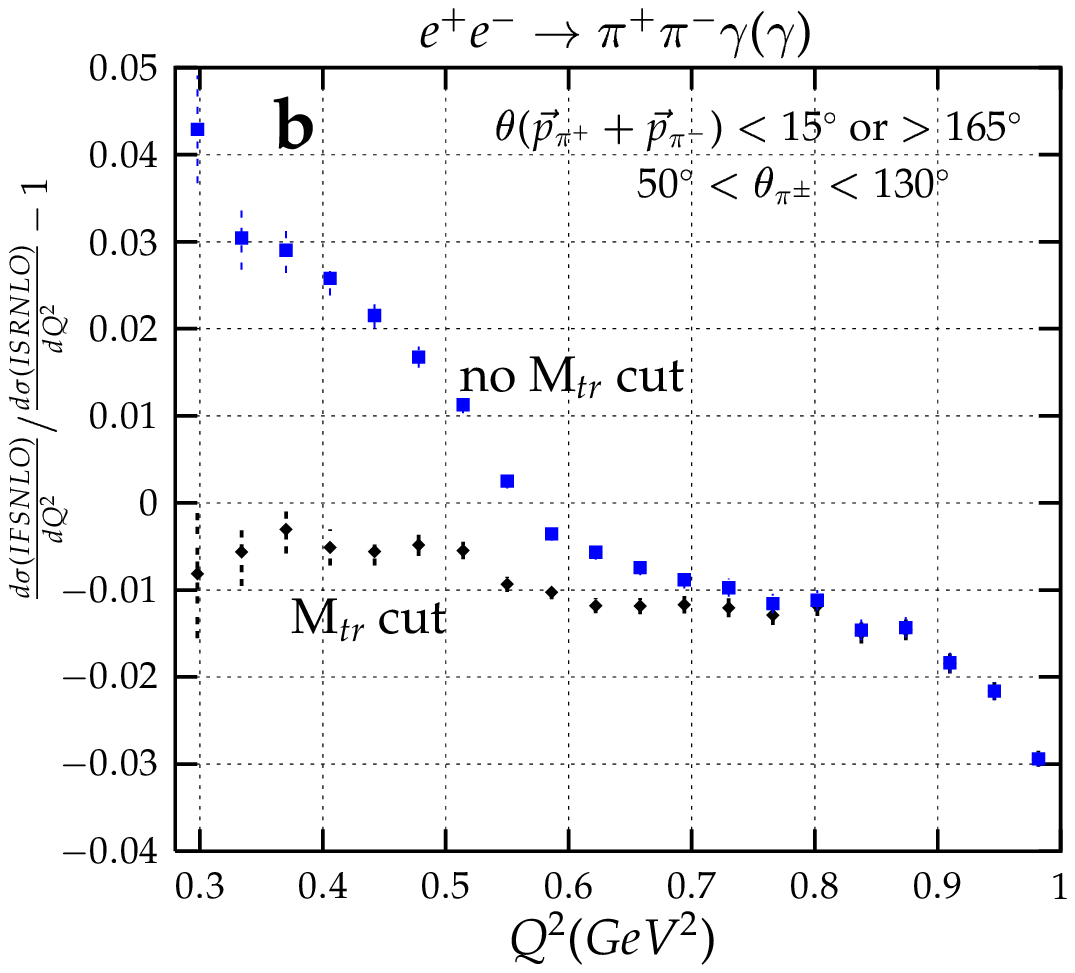,width=6.2cm,height=5.5cm}
\caption{Comparison of the \(Q^2\) differential cross sections
for \(\sqrt{s} = 1.02\)~ GeV:
IFSNLO contains the complete NLO contribution, while IFSLO has 
FSR at LO only. 
The pion and photon(s) angles are not
restricted in (a). In (b) cuts are imposed on
 the missing momentum direction and the track mass (see text for description).}
\label{fig10}
\end{center}
\end{figure}
 \begin{figure}[ht]
\begin{center}
\epsfig{file=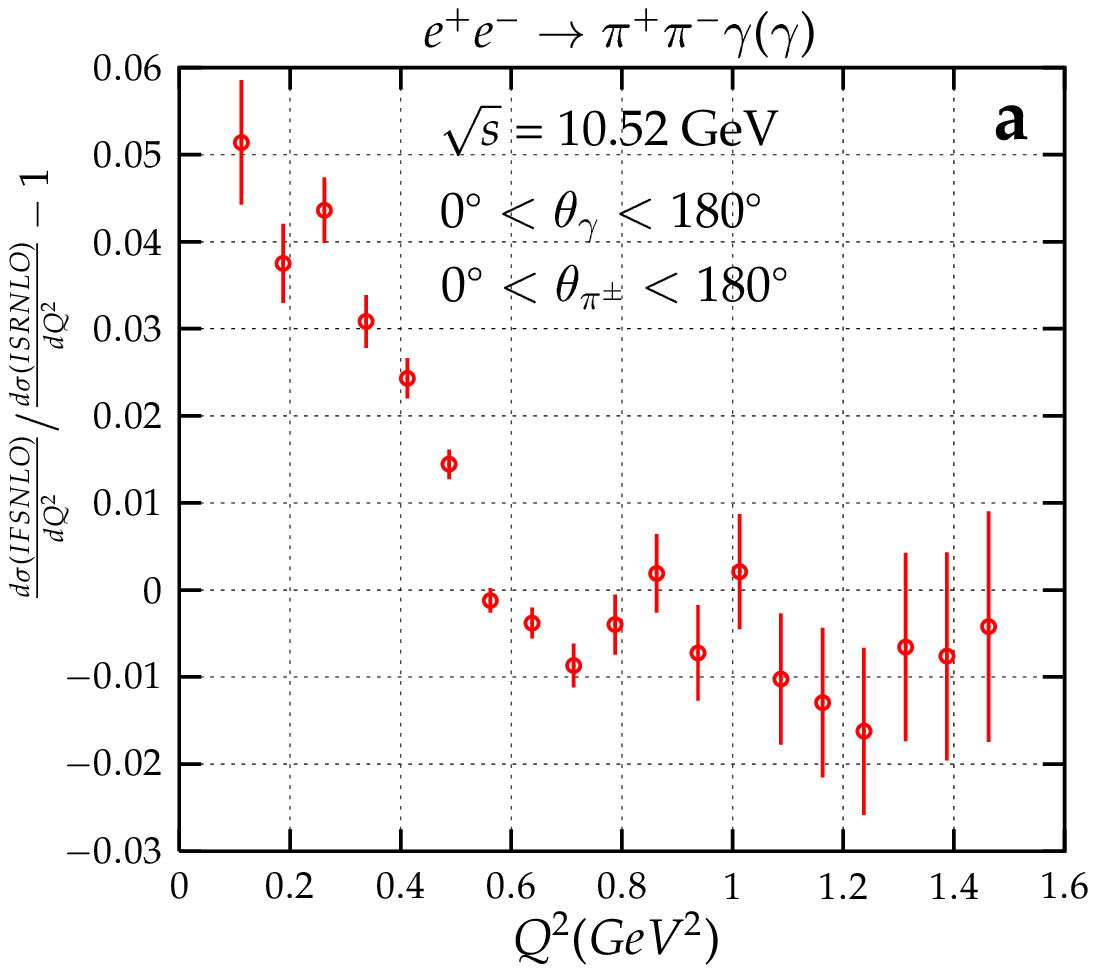,width=6.2cm,height=5.5cm}
\epsfig{file=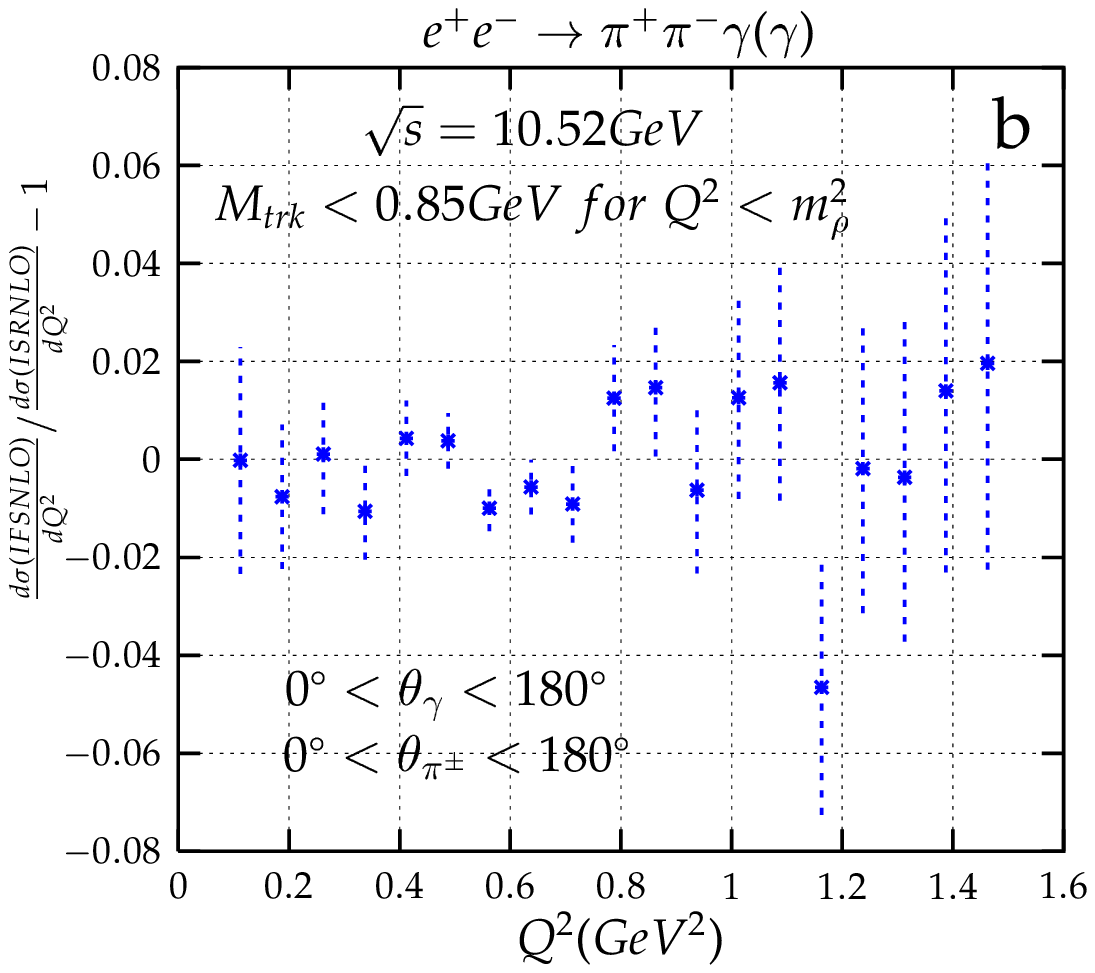,width=6.2cm,height=5.5cm}
\caption{Comparison of the \(Q^2\) differential cross sections
for \(\sqrt{s} = 10.52\)~ GeV:
IFSNLO contains the complete NLO contribution, while IFSLO has 
FSR at LO only. 
The pion and photon(s) angles are not
restricted in (a). In (b) cuts are imposed on
  the track mass for \(Q^2<m_\rho^2\) .}
\vskip -0.3cm
\label{fig11}
\end{center}
\vskip -0.5 cm
\end{figure}

 From the analysis
 of the corresponding corrections to the $e^+e^-\to\pi^+\pi^-$ process
 \cite{Schwinger:ix} in the framework of sQED, this contribution 
 is expected to be of the order of 1\%.
 However, the emission of the initial photon reduces
 the invariant mass of the $\pi^+\pi^-$ (or $\pi^+\pi^-\gamma$) system
 to the $\rho$ mass with high probability due 
to the peak of the pion form factor at the $\rho$ mass.
 As a result, 
 this contribution is strongly enhanced in the region of invariant mass of
 the $\pi^+\pi^-$ system below the $\rho$ resonance,
  as shown in Fig.\ref{fig10}a
 and Fig.\ref{fig11}a for KLOE and B--factory energies respectively.
 Suitably chosen cuts can be applied to suppress these NLO FSR 
 contributions. In Fig.\ref{fig10}b one can see that the standard KLOE
 cuts \cite{KLOE:2003,SdiFalco:Ustron}, which consist of the cuts
 on pion angles, the missing momentum angle and the track mass ($M_{tr}$),
 keep the NLO FSR contribution below 2\% with respect to the ISR cross section
 in the whole interesting region of the two--pion invariant mass.
 Similarly at B--factories, applying the track mass cut 
 only for events with $Q^2<m_\rho^2$, 
 the NLO FSR contribution is kept at a negligible level (Fig.\ref{fig11}b).

 Again, as in the case of LO FSR contributions, the main problem consists
 in the model dependence of FSR. Till now only few tests were performed
 to verify the model for FSR. However, if one aims at a measurement
 of the accuracy below 1\% such tests become indispensable.

\begin{figure}[ht]
\begin{center}
\epsfig{file=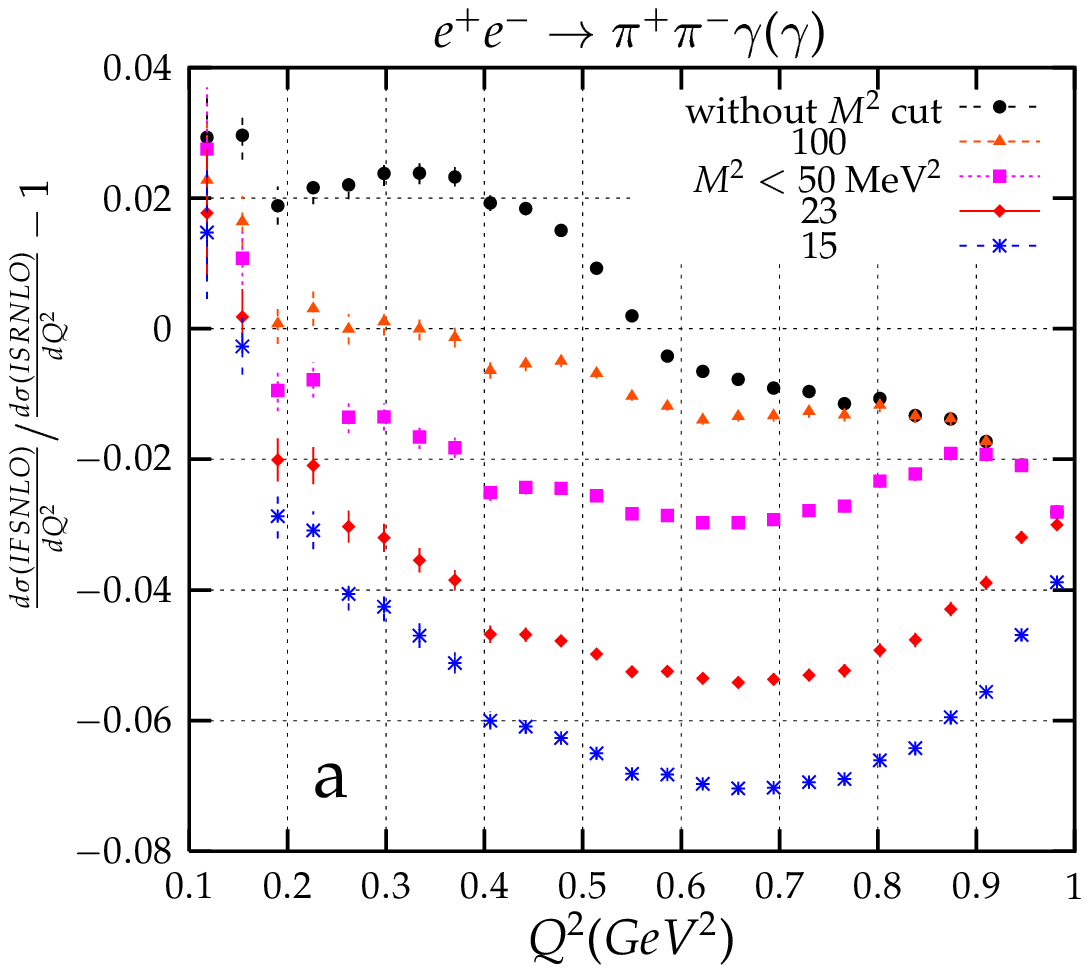,width=5.9cm,height=5.5cm}
\epsfig{file=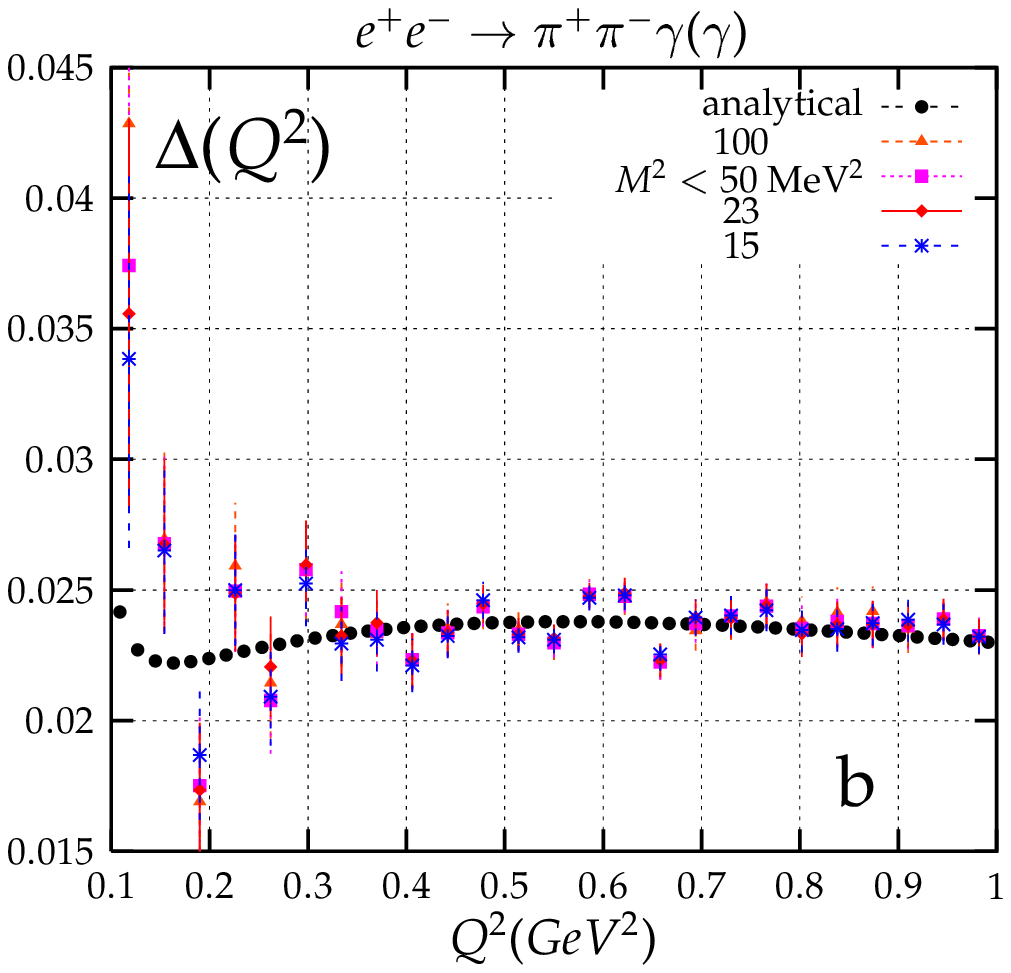,width=5.9cm,height=5.5cm}
\caption{Dependance of the relative IFSNLO contribution
 on the cut on missing invariant mass \(M^2\) (a) and radiative
 correction factor $\Delta$ extracted from MC data (see text for explanation)
  (b).}
\label{fig12}
\end{center}
\vskip -0.5 cm
\end{figure}

 In the present KLOE experimental setup, where only four momenta of the 
 pions are measured, a possibility to test the hard part of the
 NLO FSR ($\Delta$IFSNLO(H)) 
 contribution is to look at the dependence of the cross section
 on the missing invariant mass. Completely different effect of
 the cut on missing invariant 
 mass on ISR (ISRNLO) and FSR at NLO, as shown in Fig.\ref{fig12}a,
 provides a powerful tool for testing the hard part of the 
 IFSNLO contributions. A measurement of this few percent effect,
 depending on 
 the two--pion invariant mass $Q^2$ is within reach 
 of the KLOE experiment.

 The tests proposed in section \ref{secFSRLO} for the FSR at LO and the above
 tests of the hard part of the IFSNLO contributions can be further extended
 to the test of the virtual radiative corrections to the FSR of Fig.\ref{fig9}b
 following the procedure outlined below.

    The pion form factor parameters can be fitted to the data
 using the measured radiative return cross section. The cross section
 can be splited into four parts

 \bea
 \frac{d\sigma_{RR}}{dQ^2} 
  = \ {\mathrm {ISRNLO + FSRLO + \Delta IFSNLO(V+S) + \Delta IFSNLO(H)}}
 \ . 
 \eea
\noindent
 Having the form factor parameters,
 one can subtract from the data, relying on the Monte Carlo
 simulation: 
 1. the ISR contributions, which involve QED corrections only,
 2. the FSRLO contributions tested via charge asymmetries, so well known,
 and 3. the hard part of the IFSNLO corrections ($\Delta$IFSNLO(H)),
 tested as described above,
 so known with required accuracy. The left over virtual plus soft photon
 corrections to the FSR ($\Delta$IFSNLO(V+S))
 can be written in the following way

\bea
\Delta{\mathrm {IFSNLO(V+S)}} = \frac{d\sigma^{\mathrm {Born}}}{dQ^2}
\left( \ln w \cdot f + \Delta(Q^2)\right) \ ,
\eea
\noindent
where $\sigma^{\mathrm {Born}}$ stands for the LO
 $e^+e^-\to \pi^+\pi^-\gamma$ cross section with photons emitted from
 the initial state only and the $\ln w$ is the standard soft photon logarithm
 with a known function $f$ as a coefficient. 
 It is clear that one can now extract the function $\Delta(Q^2)$
 from data and
 compare it with the analytical value of the tested model.
 The function $\Delta(Q^2)$ 
 can be extracted separately in the same way for each cut on  invariant mass.
 Moreover, for self consistency of the tested model the values of 
 $\Delta(Q^2)$ obtained for each invariant mass cut should be the same.
 It provides a nontrivial check of the model,
  as the relative contribution of the
 $\Delta$IFSNLO(V+S) part to the total cross section does depend on the 
 invariant mass cut. The results of that procedure applied to the 
 Monte Carlo data obtained by PHOKHARA 3.0 are shown in Fig.\ref{fig12}b.

  Having all separate ingredients of the cross section one can 
 extract not only the pion form factor but also calculate the NLO
 corrections to the muon magnetic moment coming from the $\pi^+\pi^-\gamma$
 intermediate state.  

 \section{Conclusions}
  Extensive discussion of possible experimental tests of the model
 dependence of FSR for the $\pi^+\pi^-$ hadronic final state has been made.
  The importance of the $\pi^+\pi^-\gamma$ final state contribution
 to the muon anomalous magnetic moment has been emphasised and a method
 has been
 proposed to extract that contributions from the data using radiative
 return.
\vskip 0.1cm

 {\bf Acknowledgements:}
 The authors acknowledge J.H.K\"uhn and G. Rodrigo for fruitful collaboration
 and thank them for careful reading of the manuscript.
 The authors are grateful for the support and the kind hospitality of
 the Institut f{\"u}r Theoretische Teilchenphysik 
 of the Universit\"at Karlsruhe.



\end{document}